\documentclass[man]{apa7}

\usepackage{lipsum}
\usepackage{url}
\usepackage{graphicx}

\usepackage[american]{babel}

\usepackage{csquotes}
\usepackage[style=apa,sortcites=true,sorting=nyt,backend=biber]{biblatex}
\DeclareLanguageMapping{american}{american-apa}
\title{From Flowers to Fascism? The Cottagecore to Tradwife Pipeline on Tumblr}
\shorttitle{Cottagecore to Tradwife Pipeline?}

\author{Oliver Mel Allen, Yi Zu, Milo Z. Trujillo, Brooke Foucault Welles}
\authorsaffiliations{Network Science Institute, Northeastern University}

\authornote{
  Correspondence concerning this article should be addressed to Oliver Mel Allen at allen.ol@northeastern.edu. }

\abstract{
In this work we collected and analyzed social media posts to investigate aesthetic-based radicalization where users searching for Cottagecore content may find Tradwife content co-opted by white supremacists, white nationalists, or other far-right extremist groups. Through quantitative analysis of over 200,000 Tumblr posts and qualitative coding of about 2,500 Tumblr posts, we did not find evidence of a explicit radicalization. We found that problematic Tradwife posts found in the literature may be confined to Tradwife-only spaces, while content in the Cottagecore tag generally did not warrant extra moderation. However, we did find evidence of a mainstreaming effect in the overlap between the Tradwife and Cottagecore communities. In our qualitative analysis there was more interaction between queer and Tradwife identities than expected based on the literature, and some Tradwives even explicitly included queer people and disavowed racism in the Tradwife community on Tumblr. This could be genuine, but more likely it was an example of extremists re-branding their content and following platform norms to spread ideologies that would otherwise be rejected by Tumblr users. Additionally, through temporal analysis we observed a change in the central tags used by Tradwives in the Cottagecore tag pre- and post- 2021. Initially these posts focused on aesthetics and hobbies like baking and gardening, but post-2021 the central tags focused more on religion, traditional gender roles, and homesteading, all markers of reactionary ideals. 
}

\keywords{Social Media, Aesthetics, Gender, Radicalization, Tradwife, Cottagecore, Network Analysis, Content Analysis}

\begin{document}
\maketitle
\subsection{Introduction}
Social media can be a great way to research new hobbies and connect with others with similar interests, especially if there's a handy keyword like ``Cottagecore`` that encompasses a lifestyle you're interested in learning more about. But what happens when this keyword returns seemingly innocuous results that look similar to the content you're looking for, but also contain reactionary rhetoric? In this vein, Barbeau et al. theorize about a potential pipeline to online extremism called the ``Cottagecore to Tradwife'' pipeline (\cite{Barbeau_2022}). Some worry that through this pipeline users intending to engage in mostly innocuous Cottagecore content may end up viewing and amplifying potentially harmful and reactionary Tradwife content (\cite{Barbeau_2022},  p.23). This begs the question, how much overlap is there between Cottagecore enjoyers and reactionary Tradwives, and how do these two communities interact?

Cottagecore refers to content and content creators who post pastoral, idyllic and rural lifestyles online (\cite{O’Luanaigh_2023}). Some typical Cottagecore imagery includes rural scenery, feminine dress, and lush gardens (\cite{O’Luanaigh_2023}). Tradwives, or ‘traditional wives,’ are part of an anti-modern movement that embraces traditional gender roles in which women are the primary caretakers in their families and do not work outside the home (\cite{Zahay_2022}, \cite{O’Luanaigh_2023}). Traditional Feminism has been described as a set of values and philosophies designed to ``win over'' women and get them to reject feminism, and is often co-opted by far-right extremist communities looking for members (\cite{Proctor_2023}). 

Tradwives value the same hobbies and aesthetics seen in Cottagecore, like homemaking, feminine dress, and rural scenery, but they emphasize the role of the husband in these activities and perform them to serve a partner or be a good wife, sometimes accompanying their posts with tags like \textit{\#nationalism} and \textit{\#tradlife} (\cite{Leidig_2023}). Although Tradwife content is not inherently extremist, the literature identifies a high potential for co-opting of this content by white supremacists, white nationalists, and other far-right groups due to its proximity to both the popular Cottagecore aesthetic and reactionary ideals. Posts in online Cottagecore and Tradwife communities can serve as boundary objects \parencite{star1989institutional} that simultaneously participate in aesthetic exchange and discourse while introducing extremist content in the form of dog-whistles and suggestive imagery. Such posts can be interpreted as mundane by Cottagecore enthusiasts and may even be propagated by them, while harboring clear extremist meaning to far-right users. 

In order to discern whether and how extensively Tradwife and Cottagecore content may overlap in online spaces, potentially leading to exposure to extremist content, we performed a mixed-methods analysis of the ``Cottagecore to Tradwife pipeline'' on Tumblr. Specifically, we collected and analyzed social media posts to investigate the existence of a pipeline between Cottagecore aesthetics and Tradwife aesthetics that might lead users searching for Cottagecore content to view Tradwife content co-opted by white supremacists, white nationalists, and other far-right extremists. We investigated instances of explicit radicalization and radicalization of the environment (also known as mainstreaming). 

Through hashtag-based network analysis of over 200,000 Tumblr posts and content analysis of about 2,500 Tumblr posts, we did not find clear evidence of explicit radicalizing content in the overlap between Cottagecore and Tradwife communities on Tumblr. However, we did find evidence of Tradwife content in this overlap mainstreaming versions of white supremacist, white nationalist, or other far-right extremist ideologies. We found that Tradwife posts in the Cottagecore tag included some progressive messaging, which at first seemed promising, but may have been an effort to appeal to the norms of the Cottagecore tag. This tactic draws in users with the appeal of an accepting Tradwife community, but these users may eventually discover and amplify white supremacist, white nationalist, or far-right extremist content. Additionally,through temporal analysis we observed a change in the central tags used by Tradwives in the Cottagecore tag pre- and post- 2021. Tag use shifted from posts focused on aesthetics and hobbies like baking and gardening to more reactionary tags about homesteading and traditional gender roles. 

\section{Literature Review}
\subsection{Cottagecore Aesthetics}
Cottagecore first appeared on Tumblr in 2017 and was popular at first among the LGBTQ+ community, particularly queer women (\cite{O’Luanaigh_2023}). This stems from the belief that rural spaces are more easily ``queered'' and harkens back to the lesbian-feminist communes of the 70s (\cite{Ryan_2022}, p.169). However, the Cottagecore aesthetic does not necessarily encourage the adoption of the Cottagecore lifestyle in offline life. ``Cottagecore as a performative practice allows queer people to revel in a fictional frontier lifestyle for their own enjoyment, without concern for its actualization'' according to Ryan and Tileva (\citeyear{Ryan_2022}, p. 165). The fact that the Cottagecore lifestyle is financially unattainable for most of its aesthetic adherents is acknowledged in these spaces, and Cottagecore instead provides an opportunity for temporary escape in a relatively inexpensive and non-restrictive way (\cite{Ryan_2022}). The barrier to entry is low: one can participate in the Cottagecore aesthetic, and benefit from it, by only posting online.  

Cottagecore is not free from criticism. O’Luanaigh argues that Cottagecore aesthetics on their own lend themselves to being co-opted by extremist groups, even without Tradwives. ``The covert moral positioning and romanticisation of the past makes the Cottagecore aesthetic reactionary, appearing to favor returning to a past status quo and modes of living, and vulnerable to exploitation by reactionary actors'' (\cite{O’Luanaigh_2023}, para.5). O’Luanaigh notes that mainstream Cottagecore overlooks socio-political realities of this idealized past, like misogyny, sexism, and racism. Although posting Cottagecore images seems innocuous, the aesthetic qualities of Cottagecore posts may present an avenue for far-right actors to appropriate this content for their own gains. 

\subsection{Tradwife Aesthetics}

Like Cottagecore, Tradwife personas are also performances for social media. Leidig notes the extensive use of Pinterest by Tradwives, as well as other social media sites like YouTube and Instagram (\citeyear{Leidig_2023}). Zahay goes even further and says that ``the increasingly visual nature of mainstream platforms'' calls attention to the way that politics (in this case anti-feminist populism) is ``visually performed and displayed on our screens,'' and how alt-right women influencers take advantage of this visual medium to spread their message (\cite{Zahay_2022}, p.172). Proctor also notes that Tradwife postings are ``performances for social media audiences'' (\citeyear{Proctor_2023}, p.14). 

Unlike Cottagecore however, Tradwife content is not limited to online participation. Tradwives encourage explicit ideologies and changes in lifestyle, which can be dangerous for some women who make changes in their offline lives. According to Proctor, ``Tradwife is not just an aesthetic style or pandemic fad; it is, for many women, an identity'' (\citeyear{Proctor_2023}, p.7). Tradwife content typically ignores ramifications of extreme gender roles for women, and stories of domestic abuse are prolific on Tradwife subreddits (\cite{Leidig_2023}). Lauren Southern, a far-right woman influencer, revealed in 2024 that fully embracing a rigid, over-simplified Tradwife lifestyle led to an abusive marriage, and she knows other conservative woman influencers in similar positions (\cite{Harrington_2024}). In addition, the anti-feminist rhetoric also has implications for women who are not involved in the Tradwife lifestyle, with some Tradwives advocating against childcare, protections for women in the workplace, and college for women (\cite{Del_Valle_2023}). For some Tradwives, the only way to be happy is to actualize the Tradwife lifestyle: ``If you’re poor, it’s because your husband is failing to provide for your family. If you don’t have a husband, you haven’t been listening to God’s plan. If you’re queer, WELL, you should be miserable,'' (\cite{Petersen_2023}, para.13). 

The potential use of Tradwife content as a way to further far-right ideologies is not surprising, and the far-right has used the characterization of women as domestic figures to bring people into the movement before. According to Salice, ``The white race superiority belief system continues to construct women as domestic figures and encourages and exhorts them to identify in this role, (\citeyear{Salice_2019}, p. 5)''. This is masked as empowering by blaming feminism for women’s problems, and it pushes the idea that the natural job of a woman is to give birth to future members of the white race (\cite{Salice_2019}). White supremacists build power by exploiting vulnerabilities in mainstream culture (\cite{Belew_2022}), and the Tradwife aesthetic is one such vulnerability. Tradwives are not all anti-feminist or white nationalists, but ``the Tradwife movement appeals to and supports an infrastructure of white supremacy'' along with being inherently sexist (\cite{Proctor_2023}, p.10). For example, photos of a family gathering might not be inherently political, but when accompanied by hashtags like \textit{\#nationalism} and \textit{\#tradlife} this signals politics of the all-white nuclear family unit, and taps into the far-right’s focus on the household as a political statement (\cite{Leidig_2023}). ``The community’s hunger for the distinct boundaries of the past makes it vulnerable to far-right messaging'' and Tradwives and white nationalists share core values like more babies, myths of the west’s moral decline, and similar iconography (\cite{Leidig_2023}, p.98). Overall, the similarities between Tradwife ideals and far-right messaging are cause for concern, especially since Tradwife content is so aesthetically similar to much more mainstream Cottagecore content.

\subsection{The Hashtag Pipeline}

Both Cottagecore and Tradwife content is accessed on social media sites through hashtags and search. Ryan and Tileva (\citeyear{Ryan_2022}) note the hashtag-search feature that provides easy access to Cottagecore content, meaning those looking to seek out Cottagecore content might go to the search bar to find it. In the same way that Cottagecore posts are promoted with tags like \textit{Cottagecore}, Tradwives use hashtags like \textit{\#tradlife} and \textit{\#tradwife} to build a community online (\cite{Leidig_2023}). This use of hashtags provides a vehicle to study the overlap between these two communities on Tumblr. 

Hashtags have also been identified as a vehicle for far-right extremist groups to spread their ideology. Hashtags on Twitter that were associated with extremist ideology were found in the same tweets with hashtags associated with mainstream ideologies, a coordinated tactic known as \textit{hashtag hijacking}, and far-right groups have also been observed attempting to re-interpret or occupy specific terms (\cite{Graham2015}, \cite{Rothut_2024}). Given the mainstream popularity of Cottagecore, the aesthetic similarities between Tradwife and Cottagecore content, and the ideological similarities between Tradwife and far-right messaging, it is easy to see why Tradwife content might be a target for this hijacking, and there is some literature that shows this overlap already. For example, O'Luanaigh found a few cases of what seemed to be deliberate overlap of Tradwife and Cottagecore hashtags to increase the likelihood of Cottagecore enjoyers being shown Tradwife content (\cite{O’Luanaigh_2023}). Additionally, some TikToks posted with \textit{\#cottagecore} and \textit{\#tradwife} also included explicitly extremist or white supremacist tags (\cite{O’Luanaigh_2023}, \cite{Proctor_2023}). This extends to other benign searches, like beauty tips and wardrobe advice that might lead women to alt-right content that is aesthetically indistinguishable from innocuous content in searches (\cite{Zahay_2022}). This pipeline was defined explicitly by Barbeau et al. who posit that that users intending to engage in mostly innocuous Cottagecore content end up viewing and amplifying potentially harmful Tradwife Content (\cite{Barbeau_2022}). 

However, Grahm notes that although a user may come across extremism while searching for tweets by hashtag, that does not mean they wish to learn more about extremism or adopt extremist ideology (\cite{Graham2015}). Similarly,  Chen et al. cast doubt on the idea of a pipeline as a way to spread extremism. They found that viewers of extremist videos on YouTube already have high levels of gender and racial resentment and follow external links to the videos, not some pathway of hashtags (\cite{Chen_2023}). Winter et al. (\citeyear{Winter_2021}) also note that there is a tendency to assume that the mere existence of propaganda material equals consumption by audiences and influences on them. In this vein, even if extremist propaganda does exist in the \textit{\#tradwife} tag, it might not be consumed by or have influence on Cottagecore enjoyers.

\subsection{Tradwives, Cottagecore, and Mainstreaming}

Just because individuals might not be convinced by content they see during hashtag search, that does not mean this content has no effect on the online ecosystem. According to Rothut et al., in order to achieve their goals inconspicuously extremists adapt the presentation of their narratives to fit within mainstream content, also known as \textit{mainstreaming}. This tactic shifts broader public perception imperceptibly toward the extremes and gives extremists an increased chance of integrating their views into society (\cite{Rothut_2024}). For example, the National Socialist Movement adjusted their social media to remove Nazi imagery and replaced it with American Nationalism, and saw recruitment numbers increase (\cite{Lowell_2023}). Mainstreaming can also lead to the disassociation of radical or extremist ideology from the actors pushing that ideology (\cite{Rothut_2024}). For example,  Paul describes a ``metapolitical whiteness'' where white supremacists strategically re-frame themselves as disadvantaged and disenfranchised, disassociating whiteness from domination while calling on white supremacist histories, reviving them, and moving them back to the political center (\cite{Paul_2021}). Marwick et al. agree that ``The internet does not cause radicalization, but it helps spread extremist ideas, enables people interested in these ideas to form communities, and mainstreams conspiracy theories and distrust in institutions,'' (\cite{Marwick2022Far}).

With this framing, even if the hijacking of the Cottagecore and Tradwife tags did not radicalize individuals, it could shift public perception of feminism, women's rights, and domestic labor to the right.  At the same time, hijacking the Tradwife hashtag disassociates white nationalism and white supremacy from these ideals, replacing it with the narrative that it is a woman's choice to stay at home and live the Tradwife lifestyle. Zahay notes that ``to combat the obvious critiques of misogyny, many self-identified tradwives use feminist rhetoric to frame the movement as a choice they are making about how to live their own lives as empowered women,'' (\cite{Proctor_2023}, p. 7). 

The idea that Tradwife and adjacent white supremacist, white nationalist, and extremist content is aesthetically similar to benign influencer content, thus making it easily accessible or mainstreamed, is also supported in the literature. According to Peng et al. entertainment media like life hacks and cute animal posts are a crucial component of extreme politics landscape, and benign viral entertaining videos can boost followers and engagement with far-right outlets (\citeyear{Peng_2023}). Similarly, far-right women influencers appropriate pre-existing digital cultures like homemaking, self help, and food blogging to seem relatable and gain or retain followers (\cite{Leidig_2023}). Tradwife vloggers’ videos create the appearance of broader support by repeating aesthetics that are performed by trusted influencers, which ``primes audiences and introduces them to new pathways of extremism,'' (\cite{Zahay_2022}, p.177). Alt-right women mix personal and political content in a familiar influencer aesthetic to retain followers and appeal to wide audiences (\cite{Leidig_2023}).  The Tradwife movement broadly is not inherently extremist, but it may act as a conduit (\cite{O’Luanaigh_2023}),  and the close proximity of Tradwife ideology and white supremacist ideology on social media facilitates the exchange of ideas (\cite{Leidig_2023}). 

The benign appearance of Tradwife content may also assist in mainstreaming white supremacist ideas on Social media. Social media companies fail to recognize far-right women influencers as dangerous (\cite{Leidig_2023}), and the Tradwife movement’s ``hyper feminine aesthetic'' works to mask the ``authoritarianism of their ideology,'' (\cite{Kelly_2018}, para.8). Cottagecore content is also not exempt from this critique, and according to O’Luanaigh tech companies need to consider how content that is not explicitly violative may still contribute to radicalization processes, and how ``contextually-heavy content, such as Cottagecore content, may slip through the cracks of content moderation efforts'' (\cite{O’Luanaigh_2023}, para. 25). Thiel and McCain also note that ``White is beautiful''-style content is allowed on mainstream platforms due to the lack of overt white supremacist messaging, and on less mainstream platforms like Gab images with the same aesthetics are posted with overtly white supremacist content (\citeyear{Thiel_2022}, p.22). 

\subsection{Conclusions from the Literature}

Overall, the current literature shows that Tradwife content is rife for hijacking by white supremacists, white nationalists, and other far-right extremist groups due to its proximity to reactionary ideals and popular Cottagecore aesthetics. In this work we asked three research questions to investigate how this potential radicalization pipeline was actualized on Tumblr. First, at-scale, how much did Tradwife and Cottagecore Communities on Tumblr overlap? Second, to what extent were Tradwife posts in the Cottagecore tag explicitly radicalizing? Finally, to what extent did Tradwife posts in the Cottagecore tag mainstream reactionary ideals? 

\section{Methods}
In order to investigate the Cottagecore to Tradwife pipeline on social media we collected posts from the blogging site Tumblr using the Tumblr API\footnote{\url{https://www.tumblr.com/docs/en/api/v2}}. We chose Tumblr both because of availability of data (posts can be found all the way back to 2010) and because Cottagecore originated on Tumblr and had a well-established presence on the platform (\cite{O’Luanaigh_2023}). 

\subsection{Author Positionality}
The first, third, and last author self-identified as U.S. White
American, and the second author self-identified as Chinese/Asian. As white people the first, third, and last authors have not experienced negative consequences of and may have benefited from white supremacy, which limits their understanding of how white supremacy is experienced and expressed. Additionally, the authors did not participate in the communities of study explicitly, but some were raised in commmunities that prioritized traditional gender roles. The authors avoided bias in the methodology by building the codebook from examples found in the literature. 

\subsection{Data}
Tumblr is part blogging platform, part social network, and was launched in 2007. The platform supports multimodal blog posting including but not limited to text, audio, and video posts. It includes social media functions such as the ability to ``like'' and ``reblog'' a post and direct message other users. Tumblr also includes the ability to follow other users' blogs and provides a ``dashboard'' where these posts can be viewed (\cite{Dixon_2024}). As of 2019 Tumblr had surpassed 472 million registered accounts, and Tumblr users mostly reside in the United States, the United Kingdom, and Canada (\cite{Bianchi_2024}, \cite{Dixon_2024}). 

We collected posts by searching for specific hashtags guided by literature, which references search and tags as a vehicle for the potential pipeline (\cite{Leidig_2023}, \cite{O’Luanaigh_2023}, \cite{Zahay_2022}). 

\subsection{Quantitative Analysis}

\begin{figure}
    \caption{Number of Cottagecore posts over time on Tumblr. This matches trends in the literature and shows that most posts were made after 2018.}
    \includegraphics[width=\textwidth,height=\textheight,keepaspectratio]{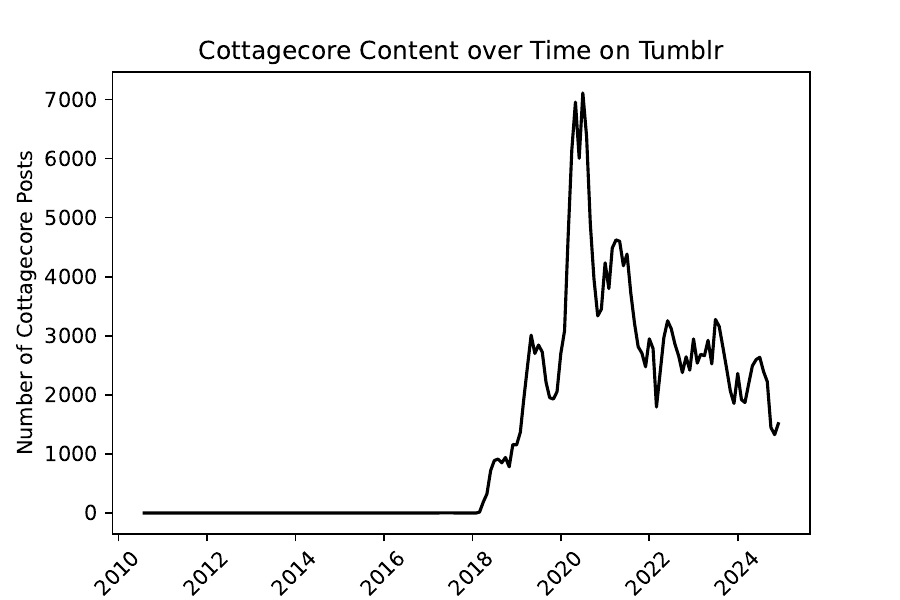}
    \label{fig:cottagecore_volume}
\end{figure}

In order to quantitatively investigate the Cottagecore to Tradwife pipeline on Tumblr we used a dataset of 223,748 original posts collected using \textit{\#cottagecore}. We chose to use posts from January 2018 to December 2024 since, as seen in figure \ref{fig:cottagecore_volume}, Cottagecore content was posted mostly after 2018. This follows trends mentioned in the literature, for example that Cottagecore posts spiked in late 2020 to early 2021 during COVID-19 lockdowns (\cite{O’Luanaigh_2023}). 

\subsubsection{Tag Co-occurence Networks}
We used tag co-occurence networks to explore the hashtag landscape in Cottagecore and Tradwife posts on Tumblr. In these networks each node is a tag and each link represents how many times two tags were used together in the same post. For example, if \textit{\#cottagecore} and \textit{\#vintage} were used in 10 posts together they would be connected with a link of weight 10. 

Often these ``raw'' networks were too large to visualize, or they contained too many irrelevant hashtags that were only posted by a few users. To resolve this we used a technique called k-core decomposition to extract the core of the tag co-occurrence network, or the tags that were posted most often and posted together multiple times. K-core decomposition trims the network by recursively eliminating nodes of a certain degree, resulting in a smaller network of more densely conected nodes and including only nodes of a certain degree or higher recursively. 

This trimmed tag co-occurence network approximates the landscape of topics frequently discussed in the tags of posts on Tumblr. Topics that are discussed together more often are closer together, and so the network represents how users might find new content from content they already consume. 

\subsubsection{Tradwife Posts in the Cottagecore Tag}
First, we examined how likely a user is to find Tradwife content when searching for \textit{\#cottagecore} by calculating the proportion of Tradwife content in this tag. We could have just counted the number of posts with \textit{\#tradwife} in our dataset, but this may have missed important content that was tagged with other similar hashtags. In order to catch more Tradwife content we used a tag co-occurence network to understand which tags were often posted alongside \textit{\#tradwife}. Then, we used K-core decomposition to trim the network and obtain a small set of candidate tags. The final set of Tradwife keywords was chosen based on the literature and included `traditional femininity', `traditional gender roles', `tradblr', `tradcat', `tradfem', `traditional', `tradwife', and `traditional wife'.  Then, using these keywords, we calculated the proportion of Tradwife posts that could be found by searching for \textit{\#Cottagecore} on Tumblr. 

\subsection{Temporal Analysis}
In order to assess mainstreaming in the overlap between Cottagecore and Tradwife communities we analyzed the overlap of the two communities at two time slices. We gathered all posts from our dataset that included at least one Tradwife hashtag from our keyword list, which included 2,494 posts made between January 2018 and December 2024 by 248 unique users. This constitutes the overlap between the Cottagecore and Tradwife communities on Tumblr (see figure \ref{fig:venn}). Based on the high proportion of Tradwife posts made during 2021 (see Figure \ref{fig:tradwife_prop}), We then split this set into posts made before January 1, 2021 and posts made after January 1, 2022. We hypothesized that the tags used pre-2021 would significantly differ from the tags used in post-2022 because most Tradwife posts in the Cottagecore tag were made during 2021. This flood of posts could have had a mainstreaming effect, where reactionary ideals were introduced into the Tradwife community, absorbed, and then exposed to those searching for Cottagecore content. Of course, there were also many other cultural and political shifts that occurred during this period, so any conclusions drawn from this analysis should consider Tumblr as a part of a larger media ecosystem. 

\subsubsection{Visualizing the Tag Landscape}
We used temporal tag co-occurence networks to visualize how the landscape of Tradwife posts in the Cottagecore tag changed over time, especially before and after 2021. We built two tag co-occurrence networks using the posts in the overlap dataset described above. We visualized these networks using Gephi (\cite{bastian2009gephi}). We trimmed the network to a manageable size for visualization (approximately 130 nodes) using weighted K-Core decomposition before visualization and used the Force Atlas layout with some adjustments to increase label visibility. The size of the node was mapped to degree and the color intensity corresponds to the proximity of that tag to the tag \textit{\#cottagecore}. The darker the node, the more times it appeared in the Cottagecore tag. 

\subsubsection{Frequency of Tags}
We also plotted the relative frequency of tags used pre-2021 and post-2022 in our overlap dataset. We used the package Shifterator (\cite{Gallagher_2021}) to calculate the proportion shift, which is the difference in relative frequencies of a word in two texts, or in our case a tag in two different time slices of Tumblr posts. If the difference is positive then the tag is relatively more common in the second slice, and if it is negative then the word is relatively more common in the first slice. A word shift graph ranks tags by this difference and plots them. This graph allowed us to see how the use of tags changed in our overlap dataset before and after 2021. 

\subsection{Qualitative Analysis}
\begin{figure}
    \caption{This Venn Diagram illustrates the population of interest. Here we collected posts with the tag \textit{\#Cottagecore}, which is the large white circle. The dark gray circle indicates Tradwife posts. The majority of our analysis concerned the light gray portion, or the overlap between these two communities, since we were interested in the Tradwife posts that one might find by searching for \textit{\#Cottagecore}.}
    \includegraphics[width=\textwidth,height=\textheight,keepaspectratio]{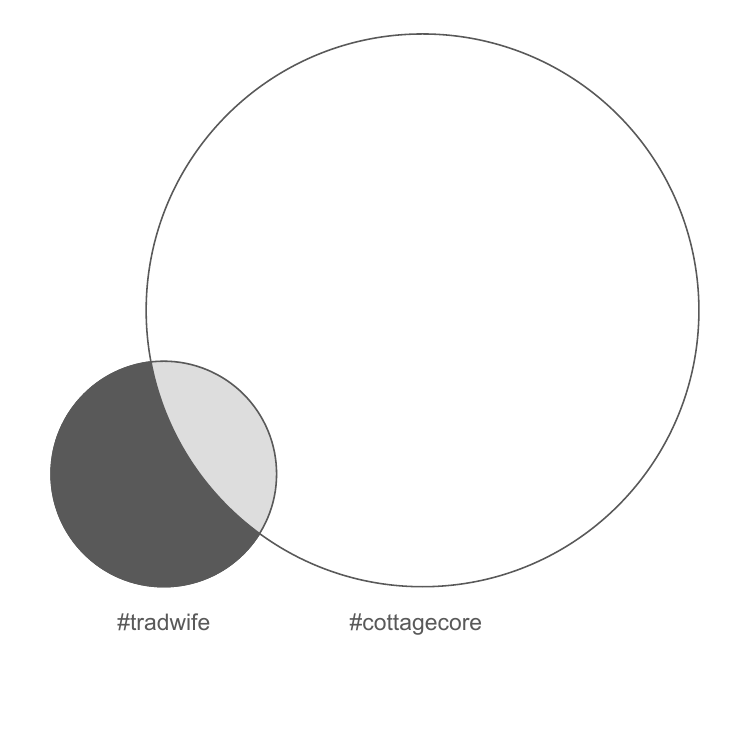}
    \label{fig:venn}
\end{figure}

In order to assess the potential for radicalization we conducted a content analysis of the posts that made up the suspected pipeline - Tradwife posts in the Cottagecore tag. We  measured what proportion of these posts contained problematic white supremacist, white nationalist, or far-right extremist content. We coded posts from the overlap dataset(see figure \ref{fig:venn}) for examples of extremist content and evidence of anti-racist or anti-homophobic sentiment. The posts were coded by two researchers and a sample was coded by both to ensure intercoder reliability. The full codebook can be seen in Figure \ref{tab:codebook} and the Fleiss' kappa scores for each code can be seen in Figure \ref{tab:resultsandfleiss}. 
 
\section{Results}

\begin{figure}
    \caption{Number of Tradwife posts in the Cottagecore tag over time on Tumblr, split between Power User and all other users. This shows a spike of posts with Tradwife-related hashtags during 2021, mostly made by Power User.}
    \includegraphics[width=\textwidth,height=\textheight,keepaspectratio]{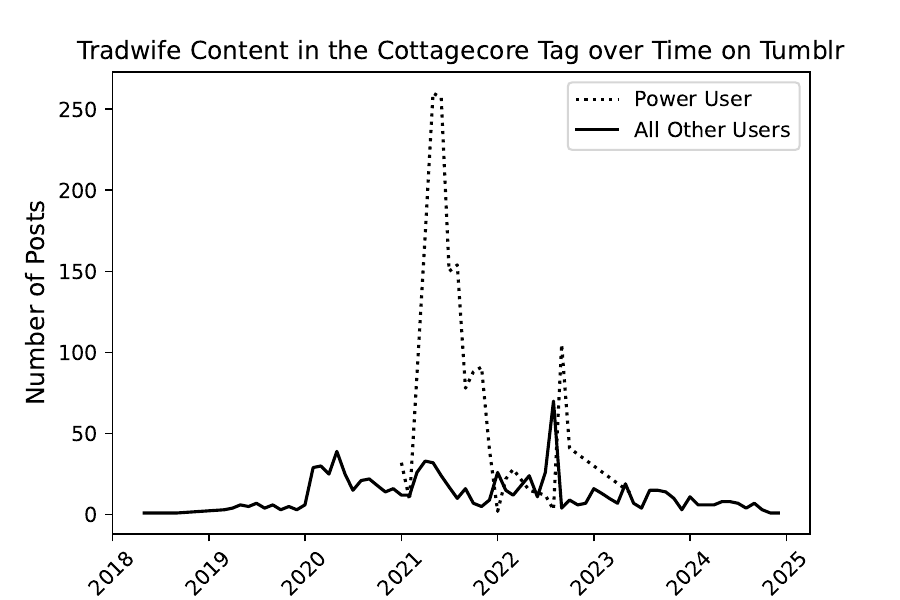}
    \label{fig:tradwife_volume}
\end{figure}

\begin{figure}
    \caption{The proportion of Tradwife posts in the Cottagecore tag over time on Tumblr.  Tradwife content was more prevalent (7 out of every 100 posts) during 2021, presumably due to Power User activity, and this continued into 2022. Otherwise there was about 1 Tradwife post per 100 posts in the Cottagecore tag.}
    \includegraphics[width=\textwidth,height=\textheight,keepaspectratio]{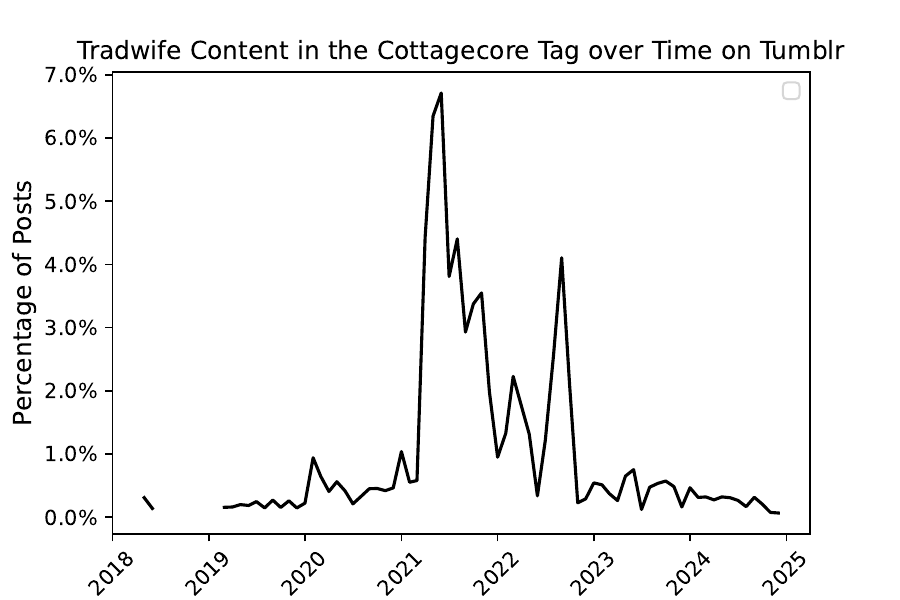}
    \label{fig:tradwife_prop}
\end{figure}

\begin{figure}
    \caption{The tag-cooccurence network created from the overlap of Tradwife posts in the \textit{\#Cottagecore} tag pre-2021. We excluded Power User from this visualization. We trimmed the network to a manageable size (~130 nodes) using weighted K-Core decomposition before visualization and used the Force Atlas layout with some adjustments to increase label visibility. The size of the node was mapped to degree and the nodes are colored according to how many times they were posted in the same post as \textit{\#Cottagecore}. Darker nodes were posted with \textit{\#Cottagecore} more often. Compared to the post-2022 visualization see a focus on aesthetic-related tags like \textit{\#flowercore} and \textit{\#forestcore} and hobbies like baking and gardening.}
    \includegraphics[width=\textwidth,height=\textheight,keepaspectratio]{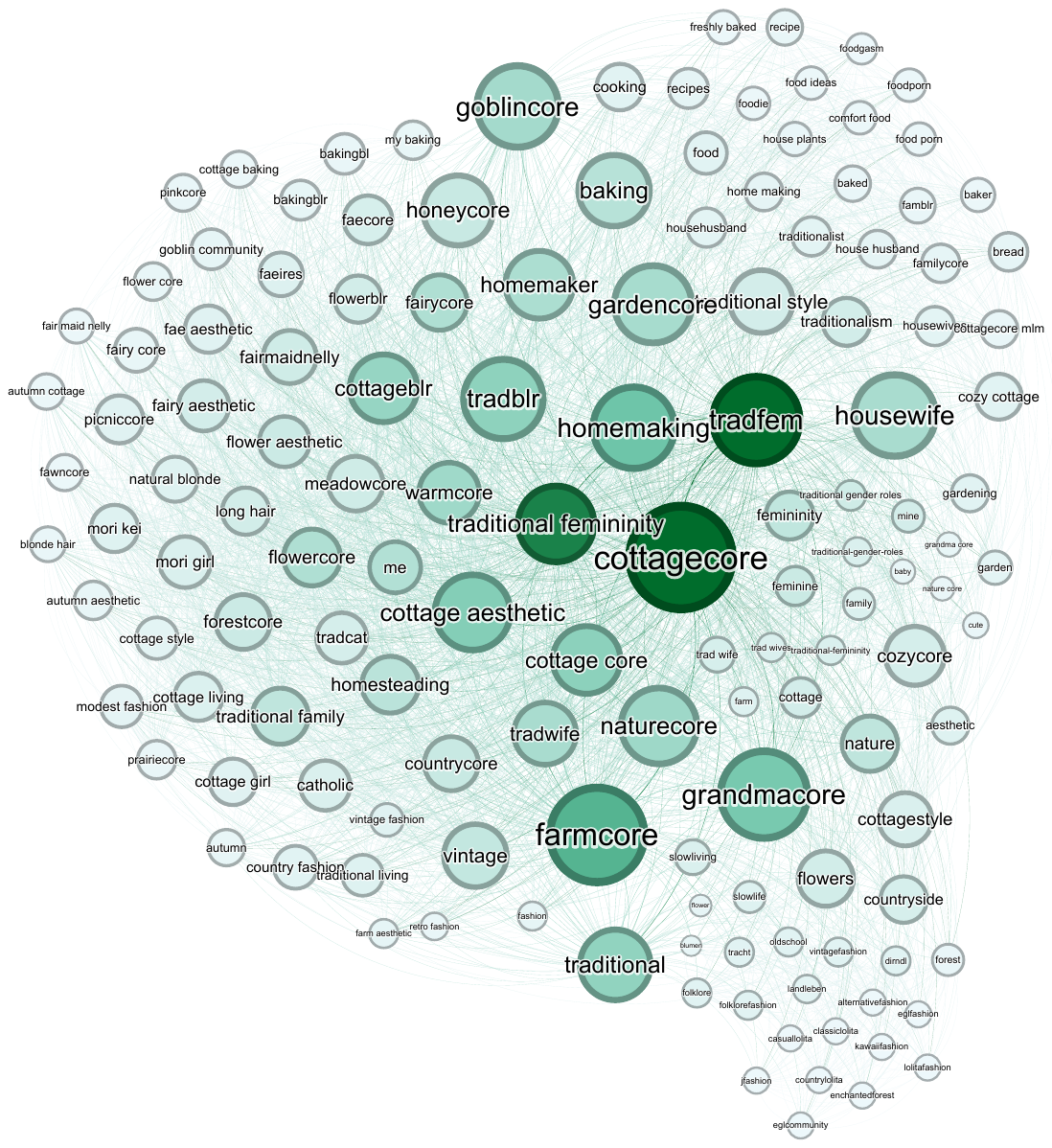}
    \label{fig:tag-tag-2021}
\end{figure}

\begin{figure}
    \caption{The tag-cooccurence network created from the overlap of Tradwife posts in the \textit{\#Cottagecore} tag post-2022. We excluded Power User from this visualization. We trimmed the network to a manageable size (~130 nodes) using weighted K-Core decomposition before visualization and used the Force Atlas layout with some adjustments to increase label visibility. The size of the node was mapped to degree and the nodes are colored according to how many times they were posted in the same post as \textit{\#Cottagecore}. Darker nodes were posted with \textit{\#Cottagecore} more often. Compared to the pre-2021 visualization we can see new tags reflecting religious values and tags reflecting a reactionary view of womanhood, for example \textit{\#traditional gender roles} and \textit{\#stay at home wife}.}
    \includegraphics[width=\textwidth,height=\textheight,keepaspectratio]{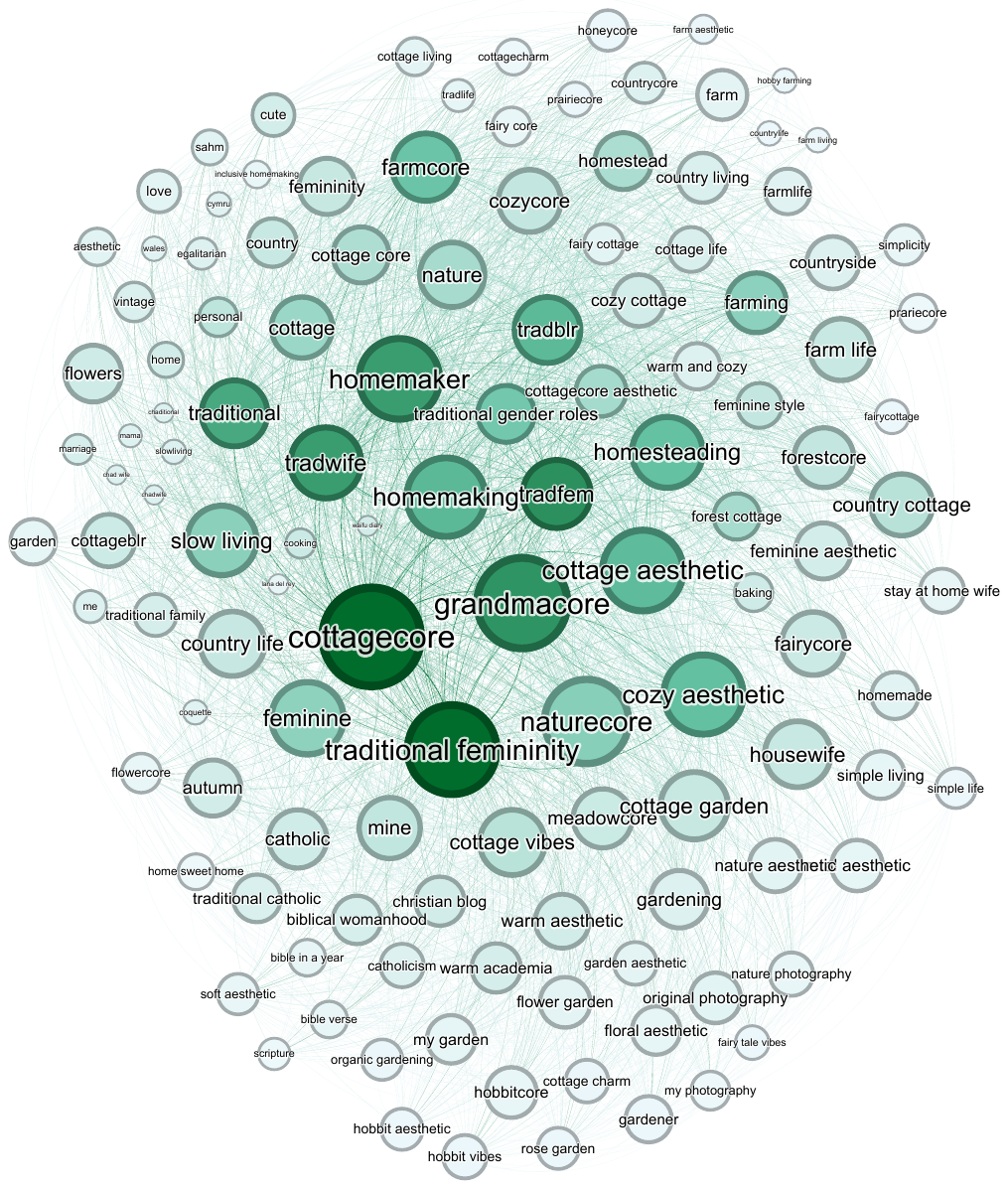}
    \label{fig:tag-tag-2022}
\end{figure}

\begin{figure}
    \caption{Proportion shift graph comparing tags used in the overlap of Tradwife posts in the \textit{\#Cottagecore} tag pre-2021 and post-2022 made with Shifterator. This excludes posts made by Power User. Any tags used to collect posts were grayed-out. The graph shows how the use of tags changed before and after 2021. Pre-2021 the highest ranked tags were focused on aesthetics, while post-2022 the highest ranked tags focused on religion, traditional gender roles, and homemaking. }
    \includegraphics[scale = 0.60,keepaspectratio]{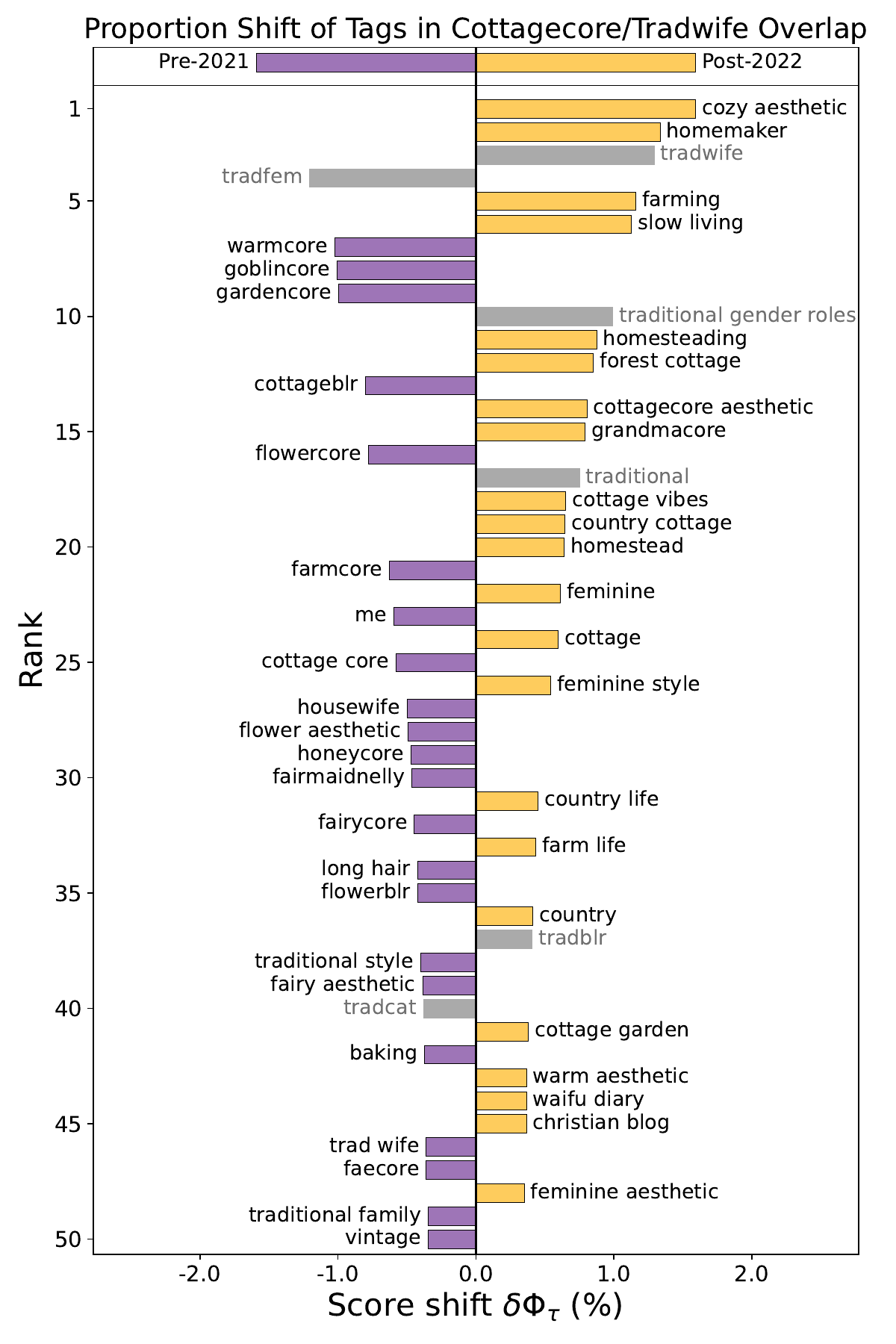}
    \label{fig:propshift}
\end{figure}

\begin{figure}
    \caption{The distribution of posts made by each blog in our overlap dataset. 100\% of blogs had at least 1 post, 10\% of blogs had at least ten posts, and 1\% of blogs had at least 70 posts. The majority of blogs had between one and ten posts, and a small percentage were much more active. There was one ``Power User`` that posted thousands of posts. }
    \includegraphics[width=\textwidth,height=\textheight,keepaspectratio]{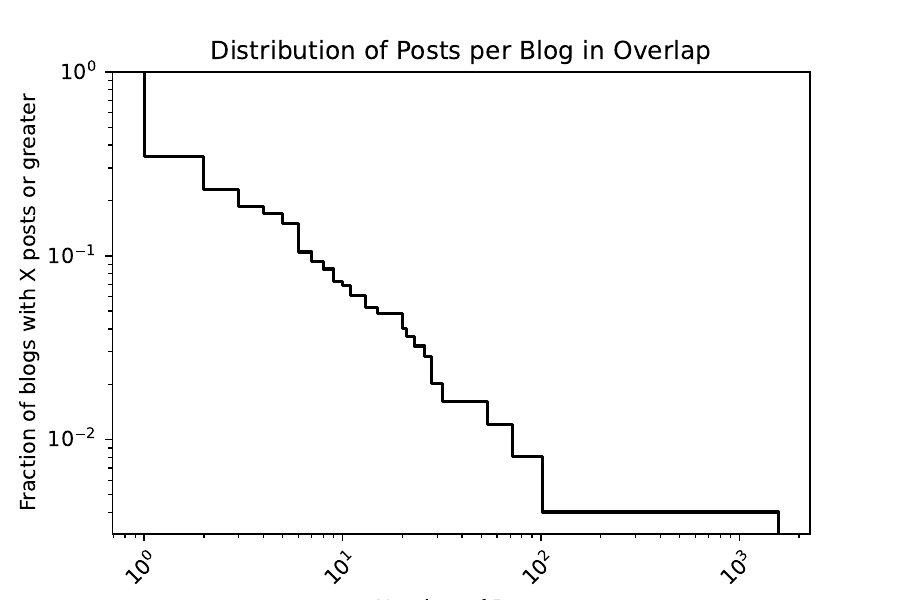}
    \label{fig:postdistribution}
\end{figure}

\begin{table}
  \caption{Codebook for Content Analysis}
  \label{tab:codebook}
  \begin{tabular}{p{0.2\linewidth}|p{0.7\linewidth}}         \toprule
  \textbf{Label}    & \textbf{Definition}  \\ \midrule
  White Supremacist Imagery      &  Includes runes, hate symbols, and any imagery that is explicitly promoting white supremacy. If you are unsure mark this as true for further examination.  If the image is used to denounce white supremacy mark 0. 

  Level one: Explicit hate symbols (mark 2)
  
  Level two: runes, Sonnenrad–icons, nordic imagery (mark 1) \\ \midrule
  White Supremacist Text       &  Includes callbacks to past values, nationalism, or other explicitly white supremacist text in the post or tags. If you are unsure mark this as true for further examination. If the post text is denouncing white supremacy mark 0 and mark 1 for POC/LGBTQ+ representation text. 
  
  Level one: explicit hate speech (mark 2)
  
  Level two: callbacks to past values, nationalism, links to suspicious telegram channels, mentions of going back to a natural state (mark 1) \\ \midrule
  POC/LGBTQ Representation Imagery       & Pictures of POC, pictures of queer relationships, queer flags, or queer symbols. POC includes anyone who is not white. We are looking for a lower bound here so if you are unsure if someone is a POC or there is not enough information mark 0.       \\ \midrule
  POC/LGBTQ Representation Text & Posts with content supporting or representing POC/LGBTQ people in the text or tags of the post. POC includes anyone who is not white. Acronyms like ``mlm'', ``wlw'' should be marked as 1. Narration in a video should be counted as text. 
      \\ \bottomrule
  \end{tabular}
\end{table}

\begin{table}
  \caption{Results and Fleiss' Kappa for each code indicating high-interrater reliability and little to no white supremacist content.}
  \label{tab:resultsandfleiss}
  \begin{tabular}{p{0.15\linewidth}|p{0.175\linewidth}|p{0.175\linewidth}|p{0.175\linewidth}|p{0.175\linewidth}}         \toprule
  \textbf{Code}    & \textbf{White Supremacist Imagery} & \textbf{White Supremacist Text}&\textbf{POC/LGBTQ Representation Imagery} & \textbf{POC/LGBTQ Representation Text}\\ \midrule
  \textbf{Fleiss' Kappa}  &  1.0 (n=668)  & 1.0 (n=668) & 0.86 (n=207) & 1.0 (n=207) \\ \midrule
  \textbf{Sampled Posts} & 0 (0\%) & 7 (0.3\%)& 61 (2.6\%)& 58 (2.5\%)\\ \midrule
  \end{tabular}
\end{table}

\subsection{RQ1: At-scale, how much did Tradwife and Cottagecore communities on Tumblr overlap?}
In our dataset we found that only 1.1\% of Cottagecore posts contained Tradwife tags (2,494 out of 223,784 posts). Looking at the proportion of Tradwife posts over time (see figure \ref{fig:tradwife_prop}), Tradwife content was more prevalent (7 out of every 100 posts) during 2021, and this continued into 2022 (4 out of every 100 posts). Otherwise there was about 1 Tradwife post per 100 posts in the Cottagecore tag with some fluctuation. One caveat is that Tumblr users have options to curate their feeds in ways that might prioritize certain posts based on engagement. However, overall, users were not likely to encounter Tradwife content when searching for Cottagecore posts on Tumblr unless they viewed hundreds of posts at a time, but users who did view many posts were likely served Tradwife content.

\subsection{RQ2: To what extent were Tradwife posts in the Cottagecore tag explicitly radicalizing? }

\subsubsection{Content Analysis}
The content analysis revealed that there were only 7 (0.3\%) Tradwife posts in the Cottagecore tag that had white supremacist or extremist content, and even these posts were not as extreme as the examples mentioned in the literature. These posts included callbacks to past values, mentions of nationalism, and links to suspicious telegram channels, and while they could be considered white supremacist when viewed from the least charitable perspective, there were no overt calls for an ethnostate, swastikas, or active recruitment (\cite{O’Luanaigh_2023}, para. 18-19). 

\subsection{ RQ3: To what extent did Tradwife posts in the Cottagecore tag mainstream reactionary ideals?}

\subsubsection{Content Analysis}

In our content analysis we found that the group of Tradwives that were posting in the Cottagecore tag were more diverse than those identified in the literature. The posts were not exclusively of white Christian straight women, with some featuring women of color and queer relationships. Overall 2.6\% of Tradwife posts in the Cottagecore tag included images of POC or LGBTQ+ people and 2.5\% included text that indicated anti-racist/anti-homophobic views. There were some examples of posts that explicitly welcomed people of color, queer people, and other marginalized groups into the Tradwife community and disavowed white supremacy. These posts presented the Tradwife community as a space for anyone, regardless of race, gender, or sexuality. There were also posts which were made by queer and POC users who were carving out a space for themselves within the Tradwife lifestyle. 

There were some posts that were ``anti-Tradwife'' which positioned the Tradwife identity in opposition to queer identity, but overall our analysis indicates that the limited Tradwife content in mainstream spaces was welcoming rather than radicalizing. This could be genuine, but previous literature suggests that this may also be a tactic to bring users looking for Cottagecore content in, lulling them into a sense of safety while working to slowly mainstream extreme ideologies. Tumblr was a hub for transgender people, at least before platform changes in 2018, so this could be a case of Tradwives specifically making space for queer identities in order to fit platform norms (\cite{Haimson_2021}). There is evidence of far-right groups accepting gay members as a concession to expand recruitment, and these efforts often also employ racism in an effort to turn white queer people against POC (\cite{Magni_2023}, \cite{Dickey2022}, \cite{Foster_2023}). Similarly, Tradwife posts that claim to welcome anyone regardless of identity may serve to bring more people into the fold while amplifying reactionary ideals about womanhood. For example, you can be queer or non-white and participate in the Tradwife community, but you must also accept those who push for anti-feminist ideas like women are meant to be in the home and serving their husbands. 

\subsubsection{Temporal Analysis}

Figure \ref{fig:postdistribution} shows that most users posted between 1 and 10 posts, and there was one user who posted thousands of posts. This is not surprising; often social media activity follows a long-tailed distribution. For example, among adult U.S. users on Twitter the most prolific 10\% create 80\% of tweets \cite{Wojcik2019}. It appears that this ``Power User'' joined Tumblr in 2021 and was responsible for the spike of posts during 2021 that we identified earlier (see figure \ref{fig:tradwife_volume}). Their content included reactionary tags like \textit{\#modesty}, \textit{\#babies}, \textit{\#modestcottagecore}, and \textit{\#traditional} posted in over 1,400 posts in our dataset. We excluded this Power User from most of our analysis to avoid attributing effects to only one user, but even without this user we still saw a qualitative difference between posts in our dataset before and after this spike of reactionary tags. 

As seen in figure \ref{fig:propshift}, as we hypothesized, the hashtag landscape changed significantly before and after 2021. Pre-2021 Tradwife posts in the Cottagecore space were mainly concerned with a few Tradwife-specific tags like \textit{\#tradfem} and \textit{\#tradcat} and other aesthetic-related tags like \textit{\#flowercore}, \textit{\#forestcore}, and \textit{\#goblincore}.  Post-2022 posts in the same space shifted to focus more on homemaking and traditional roles, for example\textit{ \#traditional gender roles}, \textit{\#homesteading} and \textit{\#christian blog}. 

In addition, this same trend is supported by the visualizations of the tag-cooccurence networks. Although the visualization is less precise, we can still observe which tags appeared close to \textit{\#cottagecore} and which appeared peripheral in the network. In the pre-2021 network aesthetic-related tags like \textit{\#flowercore}, \textit{\#forestcore}, and \textit{\#goblincore}, made up much of the network, along with hobbies like baking and gardening. Post-2022 tags focused on motherhood, religion, and homesteading, for example \textit{ \#traditional gender roles}, \textit{\#biblical womanhood} and \textit{\#stay at home wife} appeared larger and closer to \textit{\#cottagecore}.

This shows that although there is not much evidence of explicit radicalization in Tradwife posts in the Cottagecore tag, the landscape of this overlap changed pre-and post-2021. Our analysis shows a change from Internet-based aesthetics like \textit{\#fairycore} to a focus on religion, traditional roles of women, and the reactionary fantasy of an idyllic, pastoral lifestyle often evoked by eco-fascists and extremists (\cite{O’Luanaigh_2023}, para. 23-26). This slower, less explicit shift in values and topics indicates that even though those searching for Cottagecore content were very unlikely to be exposed to explicit white supremacist posts, they could have been exposed to Tradwife posts in the Cottagecore tag that normalized a reactionary view of womanhood and femininity. Additionally, as seen in the literature these posts lie in a moderation gray area due to their non-explicit nature, making it difficult to protect those searching for Cottagecore content with moderation alone. 

\subsection{Limitations}

This research is limited to posts on Tumblr, but this is just one small part of the larger social media ecosystem. Tradwife, Cottagecore, and far-right extremist content can be found on many social media sites like Pinterest, TikTok, and Instagram, which all have different dynamics and users, and should be investigated to gain a cross-platform understanding of this phenomena. The concept of mainstreaming is, by definition, a gradual and imperceptible shift, so any conclusions drawn from this research should take into account not only the broader social media landscape, but also the effects of offline news and events. In addition, this research focused on a hashtag pathway, which is only one mode of interaction on Tumblr, leaving out the dynamics of reblogs, following behavior, and private messaging. Finally, the Tumblr API documentation\footnote{\url{https://www.tumblr.com/docs/en/api/v2}} does not specify exactly how posts are collected, and although we gathered a significant number of posts, there is no way to guarantee the representativeness of our data. 

\section{Conclusion}
In this work we asked three research questions to investigate how a potential radicalization pipeline from Cottagecore to Tradwife communities was actualized on Tumblr. First, at-scale, how much do Tradwife and Cottagecore Communities on Tumblr overlap? Second, to what extent are Tradwife posts in the Cottagecore tag explicitly radicalizing? Finally, to what extent to Tradwife posts in the Cottagecore tag mainstream reactionary ideals? 

We found that Tradwife posts made up only 1.1\% of posts in the Cottagecore tag, and in these posts there were no explicit calls to radicalization or declarations of white supremacy. Overall, our analysis does not support an explicit pipeline of radicalization, and problematic Tradwife posts found in the literature may be confined to Tradwife-only spaces. However, we did find evidence that white supremacist, white nationalist, and other far-right extremist ideologies were mainstreamed in Tradwife posts in the Cottagecore tag. In our content analysis there was more overlap between queer and Tradwife identities than expected based on the literature, and some Tradwives even explicitly included queer people and disavowed racism in the Tradwife community on Tumblr. This could be genuine, but the literature suggests this may have been an attempt to re-brand content and follow platform norms to spread ideologies that would otherwise be rejected by Tumblr users. Additionally, through temporal analysis we observed a change in the tags used by Tradwives in the Cottagecore tag pre- and post- 2021. Initially these posts focused on aesthetics and hobbies like baking and gardening. Post-2021 the central tags focused more on religion, traditional gender roles, and homesteading, all markers of reactionary ideals. This shift indicates that even though those looking for Cottagecore content were very unlikely to be exposed to explicit white supremacist posts, the Tradwife Posts in the Cottagecore tag mainstreamed reactionary ideals about womanhood while potentially evading moderation due to their non-explicit nature. 

Future research could replicate this study on other social media platforms such as Pinterest, TikTok, or YouTube, or investigate other subjects like lifestyle vlogging or nutrition. Additionally, we did find one post linking out to a Telegram channel that contained white supremacist content according to our codebook, so it is possible that users are being recruited to radical communities through other mechanisms. This study focused on hashtags, but future work could investigate reblogs, following relationships, or recruitment via private messaging. Further research could also focus on why POC and queer people post in the Tradwife tag, and attempt to untangle the complex relationship between queer identity, racial identity, and Tradwife identity on Tumblr. 

\printbibliography


\end{document}